\documentclass[usenatbib]{mn2e}
\usepackage{graphicx}
\usepackage{epsfig,psfrag}
\usepackage{natbib}
\usepackage{pslatex}
\usepackage{graphicx}
\usepackage{gensymb}

\def\hMsun{h^{-1} \; M_\odot}
\def\hMpc{h^{-1} \; {\rm Mpc}}

\title[HOD of WISE Quasars]{The Halo Occupation Distribution of Obscured Quasars: Revisiting the Unification Model} 
\author[Mitra et al.]{\parbox{17cm}{Kaustav Mitra$^{1}$, Suchetana Chatterjee$^{1}$, Michael A. DiPompeo$^{2}$, Adam D.\ Myers$^{3}$ and Zheng Zheng$^{4}$}
\vspace{0.2cm}\\
$^{1}${Department of Physics, Presidency University, Kolkata, 700073, India}\\
$^{2}${Department of Physics and Astronomy, Dartmouth College,
			Hanover, NH 03755, USA}\\
$^{3}${Department of Physics and Astronomy, University of Wyoming,
			Laramie, WY 82071, USA}\\
$^{4}${Department of Physics and Astronomy, University of Utah,
			Salt Lake City, UT 84112, USA}\\}

\begin{document}
\maketitle

\begin{abstract}
We model the projected angular two-point correlation function (2PCF) of obscured and unobscured quasars selected using the Wide-field Infrared Survey Explorer (WISE), at a median redshift of $z \sim 1$ using a five parameter Halo Occupation Distribution (HOD) parameterization, derived from a cosmological hydrodynamic simulation by Chatterjee et al. The HOD parameterization was previously used to model the 2PCF of optically selected quasars and X-ray bright active galactic nuclei (AGN) at $z \sim 1$. The current work shows that a single HOD parameterization can be used to model the population of different kinds of AGN in dark matter halos suggesting the universality of the relationship between AGN and their host dark matter halos. Our results show that the median halo mass of central quasar hosts increases from optically selected ($4.1^{+0.3}_{-0.4} \times 10^{12} \; h^{-1} \; \mathrm{M_{\sun}}$) and infra-red (IR) bright unobscured populations ($6.3^{+6.2}_{-2.3} \times 10^{12} \; h^{-1} \; \mathrm{M_{\sun}}$) to obscured quasars ($10.0^{+2.6}_{-3.7} \times 10^{12} \; h^{-1} \; \mathrm{M_{\sun}}$), signifying an increase in the degree of clustering. The projected satellite fractions also increase from optically bright to obscured quasars and tend to disfavor a simple `orientation only' theory of active galactic nuclei unification. Our results also show that future measurements of the small scale clustering of obscured quasars can constrain current theories of galaxy evolution where quasars evolve from an IR- bright obscured phase to the optically bright unobscured phase.  
\end{abstract}

\section{Introduction}

There is now a great deal of evidence linking galaxy evolution to the growth of supermassive black holes \citep[SMBH; e.g.,][]{richstoneetal98, gebhardtetal00, m&f01, tremaineetal02, grahametal11}. The cold dark matter paradigm of galaxy formation implies that galaxies form in the potential wells of massive dark matter (DM) halos \citep[e.g.,][]{w&r78, w&f91, kauffmannetal93, nfw95, m&w96, kauffmannetal99, hopkinsetal10, c&w13, conselice14, shankaretal15}. So, a complete assessment of galaxy evolution requires an understanding of the connection between the growth and formation of SMBH and the dark matter halos they inhabit.

Galaxies that emit particularly strongly from the region near the SMBH that they harbor are called active galactic nuclei (AGN). AGN have been used to study the interplay between dark matter halos, and the galaxies and SMBHs they host, which is 
often referred to as ``AGN/SMBH co-evolution''  \citep[e.g.,][]{k&h00, w&l03, marconietal04, cattaneoetal06, crotonetal06, hopkinsetal06, lapietal06, shankaretal04, dimatteoetal08, b&s09,  volonterietal11, c&w13, caplaretal15, oogietal16}. 

A key observational probe of the relation between SMBHs and their host DM halos is AGN clustering, which is frequently measured via the two-point-correlation function \citep[2PCF; e.g.,][]{arp70}. Clustering measurements of different types of AGN have been carried out by several groups employing data from multiple surveys in the optical waveband \citep[e.g.,][]{croometal04, porcianietal04, croometal05, gillietal05, myersetal06, myersetal07a, myersetal07b, coiletal07, shenetal07, wakeetal08, shenetal09, rossetal09, coiletal09, hickoxetal09, hickoxetal11, allevatoetal11, donosoetal10, krumpeetal10, cappellutietal12, whiteetal12, shenetal12a, krumpeetal12, mountrichasetal13, koutoulidisetal13, krumpeetal15, eftekharzadehetal15, eftekharzadehetal17}.

The majority of these studies involve measurement of the 2PCF of a certain kind of AGN, namely optically bright quasars. Due to their high luminosity, quasars are detected to high redshifts \citep[as high as $z \sim 7$, e.g.,][]{mortlocketal11}, making them powerful probes of structure formation over a wide redshift range. In addition, the large sample sizes of quasars and the availability of reliable redshifts make them excellent candidates for studying how SMBHs co-evolve with cosmic structure. However, quasars have broad spectral-energy distributions and quasar emission at different wavelengths may be characteristic of quite different physical processes in the accretion disc and adjacent structures surrounding the central engine. Studies of quasar clustering have therefore moved beyond the optical waveband spanning the entire electromagnetic spectrum to test how large scale structures influence the properties of quasars \citep[e.g.,][]{shenetal09, donosoetal10, hickoxetal11, dipompeoetal14, mendezetal16, dipompeoetal16}. 

In this work, we employ the halo occupation distribution (HOD) formalism \citep[e.g.,][] {m&f00, seljak00, b&w02, zhengetal05, z&w07, wakeetal08, shenetal10, miyajietal11, starikovaetal11, allevatoetal11, richardsonetal12, k&o12, shenetal12a, richardsonetal13, allevatoetal14, c&s15b} to derive the host dark matter halo properties of the quasars studied by \citet[][hereafter D16]{dipompeoetal16}. 

The HOD technique has been successfully used in the context of galaxy evolution in the recent past \citep[e.g.,][]{zehavietal05, zhengetal05, z&w07, zhengetal07}. Currently, large multi-wavelength datasets of AGN have provided us the tools to carry out the HOD analysis of AGN/quasar clustering in a statistically robust manner. Recently (\citealt{richardsonetal12}; R12 hereafter and \citealt{richardsonetal13}; R13 hereafter) use optically selected quasars and X-ray bright AGN at $z \sim 1$ to perform a comparison study of the HOD using the measured 2PCF of these two classes of AGN. 

The results show that a universal parametrization of the AGN HOD is applicable for these two classes of AGN suggesting a universality in the relationship between AGN and their host dark matter halos. The results favor a scenario in which SMBH are believed to evolve from a bright quasar phase to an X-ray phase to a radio-loud phase along with the growth and evolution of their host dark matter halos. This scenario was proposed by \citet{hickoxetal09} using multi-wavelength samples of low redshift AGN. The HOD technique is hence emerging as a successful tool to study quasar/AGN co-evolution in the way it allowed us to understand galaxy evolution with large scale structure in the Universe.

\begin{figure*}
\begin{center}
\begin{tabular}{c}
        \resizebox{8cm}{!}{\includegraphics{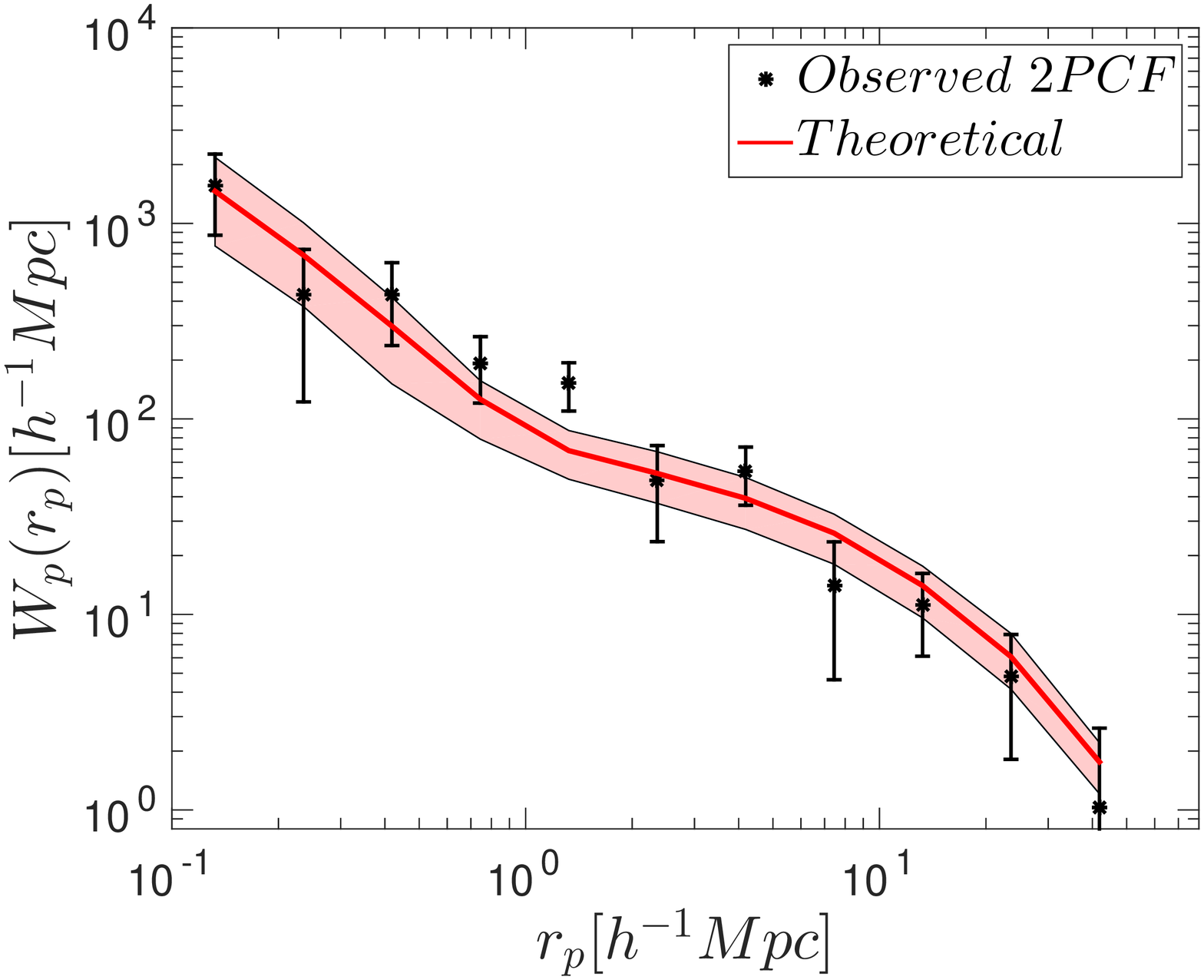}}
        \resizebox{8cm}{!}{\includegraphics{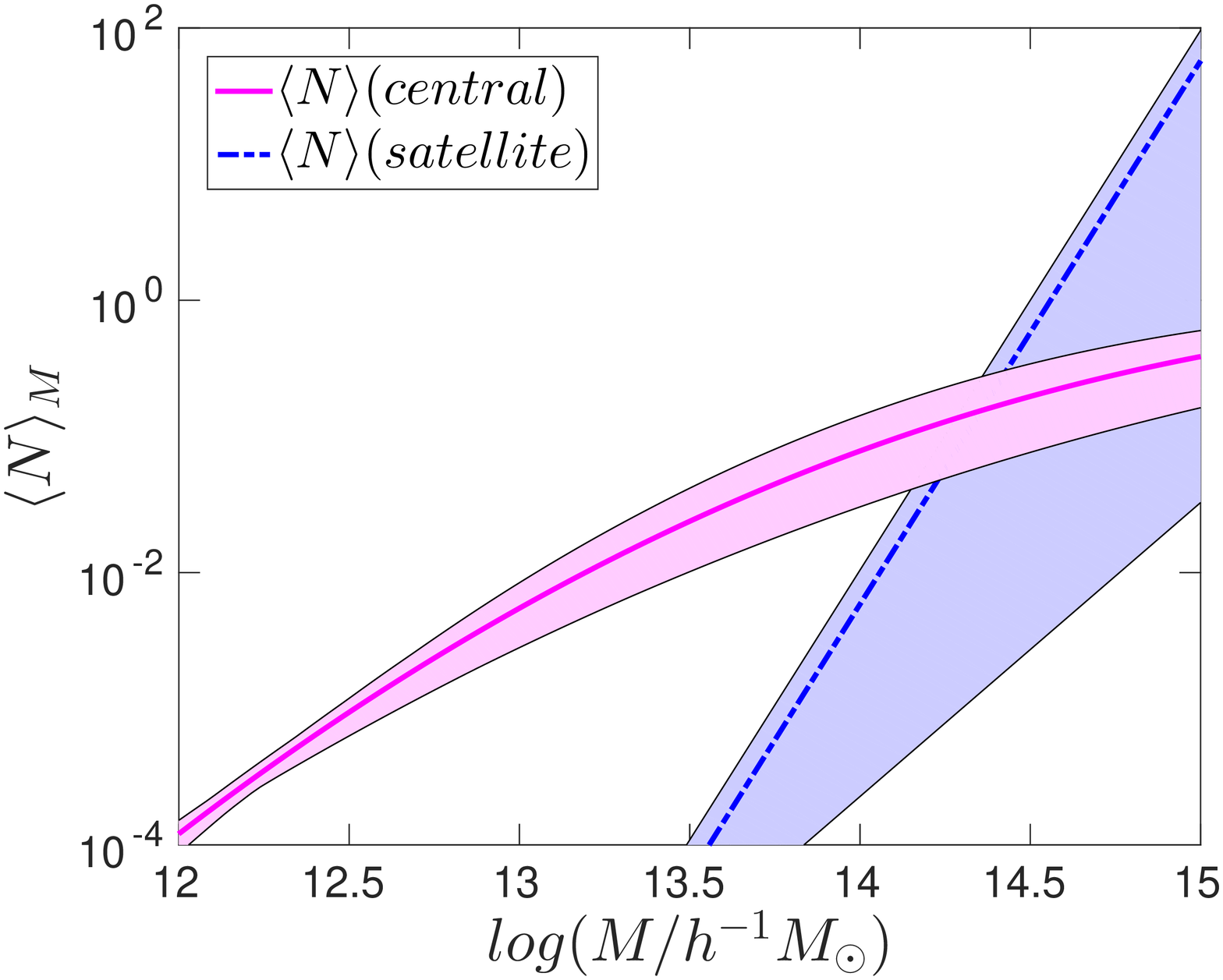}}\\
         \resizebox{8cm}{!}{\includegraphics{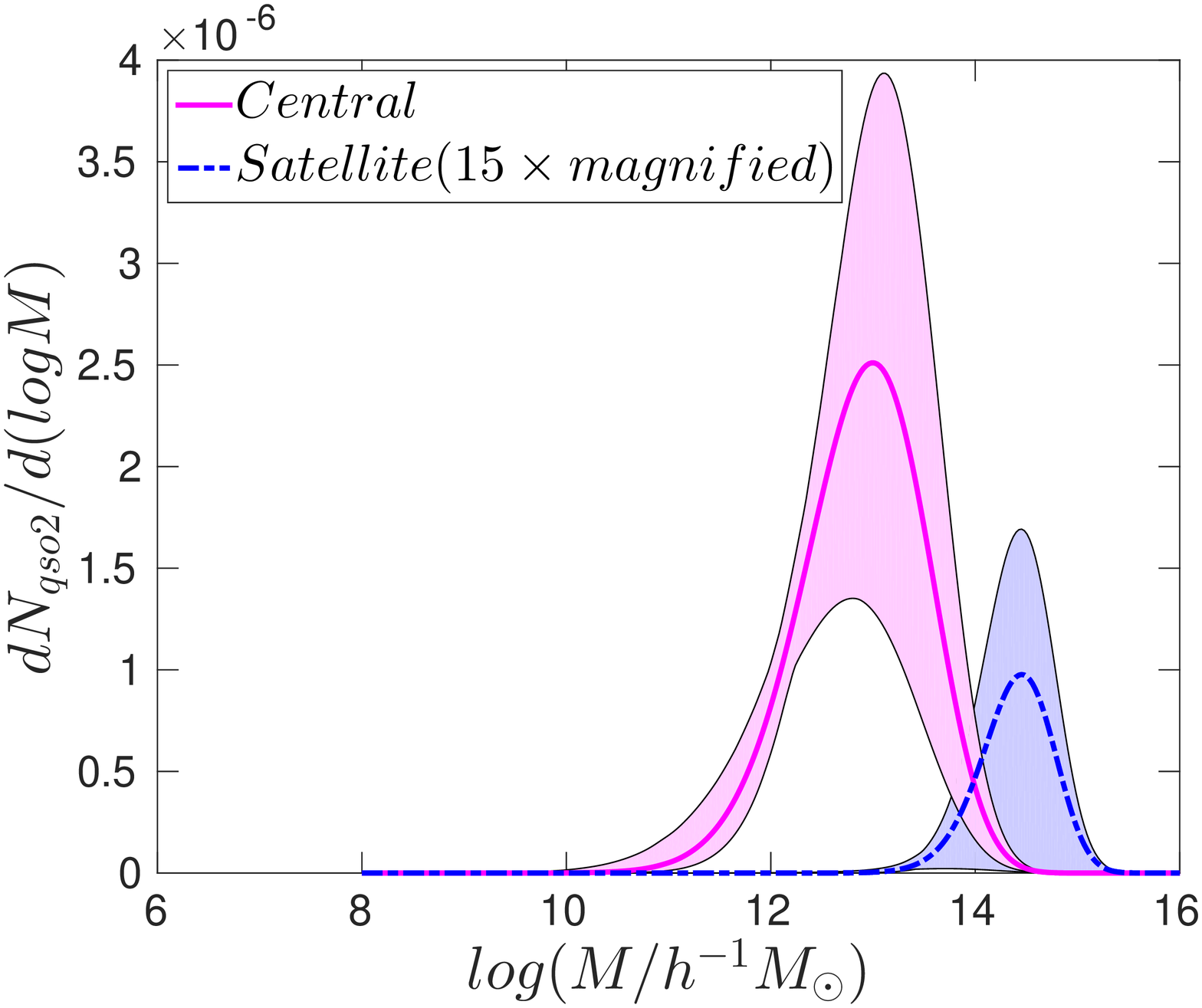}}
         \resizebox{8cm}{!}{\includegraphics{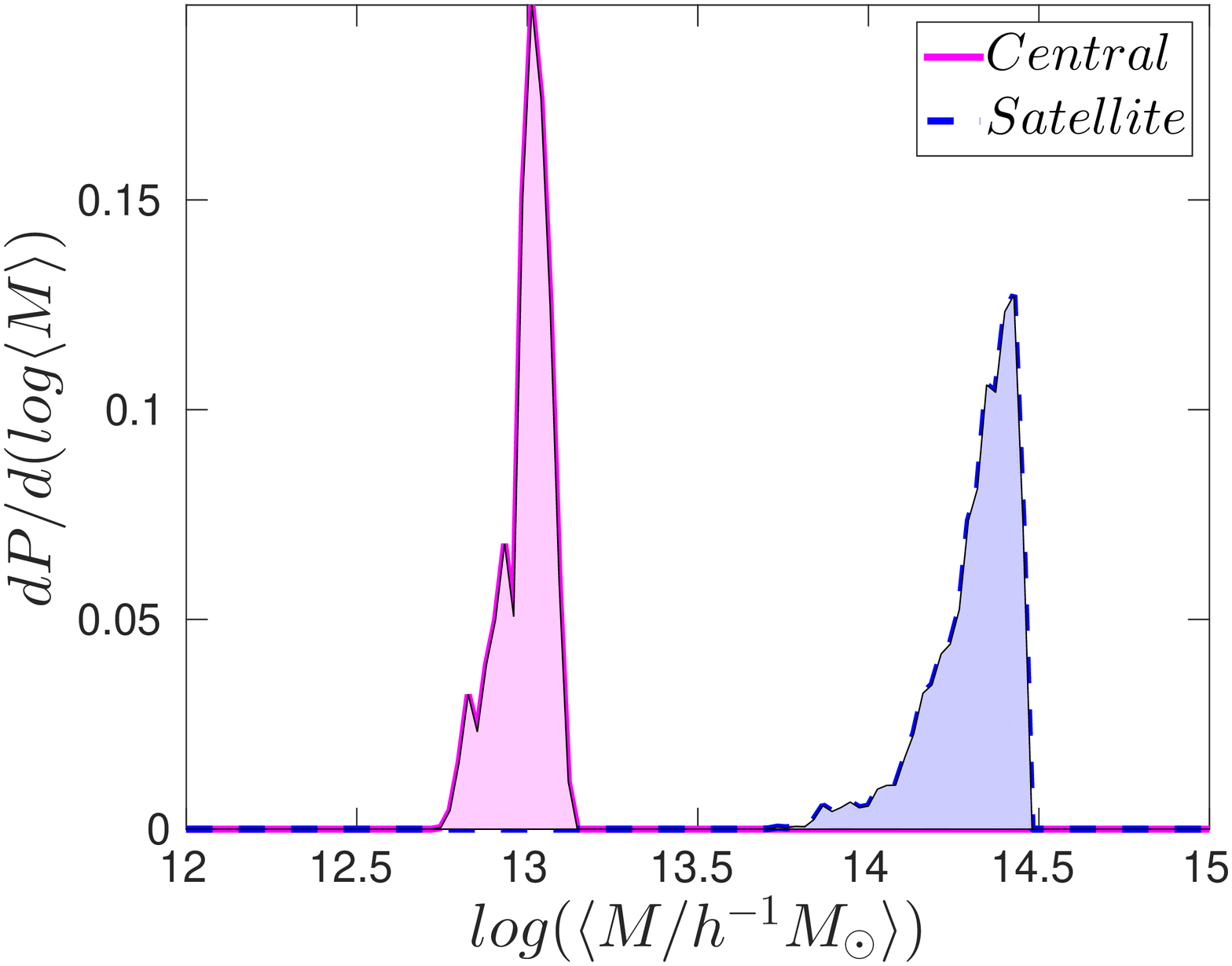}}\\
          \end{tabular}
 \caption{{\bf Top Left} : The projected 2PCF of the WISE-selected obscured quasar sample (median redshift $0.9$) from \citet{dipompeoetal16}. The red solid line and and the red shaded region correspond to the best-fit model and the error on it respectively. {\bf Top Right} : The Mean Occupation Function (MOF) constructed from the best-fit parameters. The magenta solid line and the blue dashed line, with corresponding errors, are the MOFs of the central and satellite quasars respectively. {\bf Bottom Left} : The distribution of central (magenta solid line, with shaded errors) and satellite (blue dashed line, with shaded errors, scaled by a factor of fifteen for visualization purpose) obscured quasars in dark matter halos as a function of halo mass. {\bf Bottom Right} : The probability distributions of the median mass scales of central (magenta solid line) and satellite (blue dashed line) obscured quasars that produce a model 2PCF consistent with the observed data (see discussions in \S 4). }
\end{center}
\end{figure*}

Recently D16 measured the 2PCF of $z \sim 1.0$ quasars, selected using infrared imaging from the Wide-field Infrared Survey Explorer \citep[WISE;][]{wright10}, and characterized as obscured or unobscured using an optical-IR color cut with optical imaging data \citep{hickoxetal07} from Data Release 8 (DR8) of the Sloan Digital Sky Survey \citep[SDSS;][]{york00}. Our work is in the vein of R12 and R13 where we try to test for the universality of the AGN HOD using obscured quasars and assess the role of the obscured phase in light of our previous work with optical and X-ray bright sample. 

We compare the HOD properties of the WISE-selected quasars with the optical sample to test for similarities and$/$or differences in the large scale (and intra-halo) environments of these two classes of quasars. According to the simplest AGN unification theory, the central SMBH and accretion disk of a quasar are surrounded by an optically thick dusty torus, and the obscuration of the central broad-line-region and the accretion disk by the torus occurs at certain inclination angles as the torus intercepts the line of sight \citep{u&p95}. In this paradigm, it is expected that different population of quasars would have identical host halo properties as the distinction is only an orientational effect. Our work is directed toward examining the validity of this prediction via a robust, HOD-based approach.

The paper is organized as follows. In $\S2$ and $\S3$, we briefly describe our data sets, the parameterization of the quasar HOD, and the theoretical modeling of the 2PCF. We present the results of our HOD modeling in $\S4$. Finally, we discuss the implications of our results and summarize them in $\S5$. Throughout this work we assume a spatially flat, $\Lambda$CDM cosmology \citep{spergeletal07}: $\Omega_{m}=0.26$, $\Omega_{\Lambda}=0.74$, $\Omega_{b}=0.0435$, $n_{s}=0.96$, $\sigma_{8}=0.78$, and $h=0.71$. We quote all distances in comoving $\hMpc$ and masses in units of $\hMsun$ unless otherwise stated.

%
\begin{figure*}
\begin{center}
\begin{tabular}{c}
        \resizebox{8cm}{!}{\includegraphics{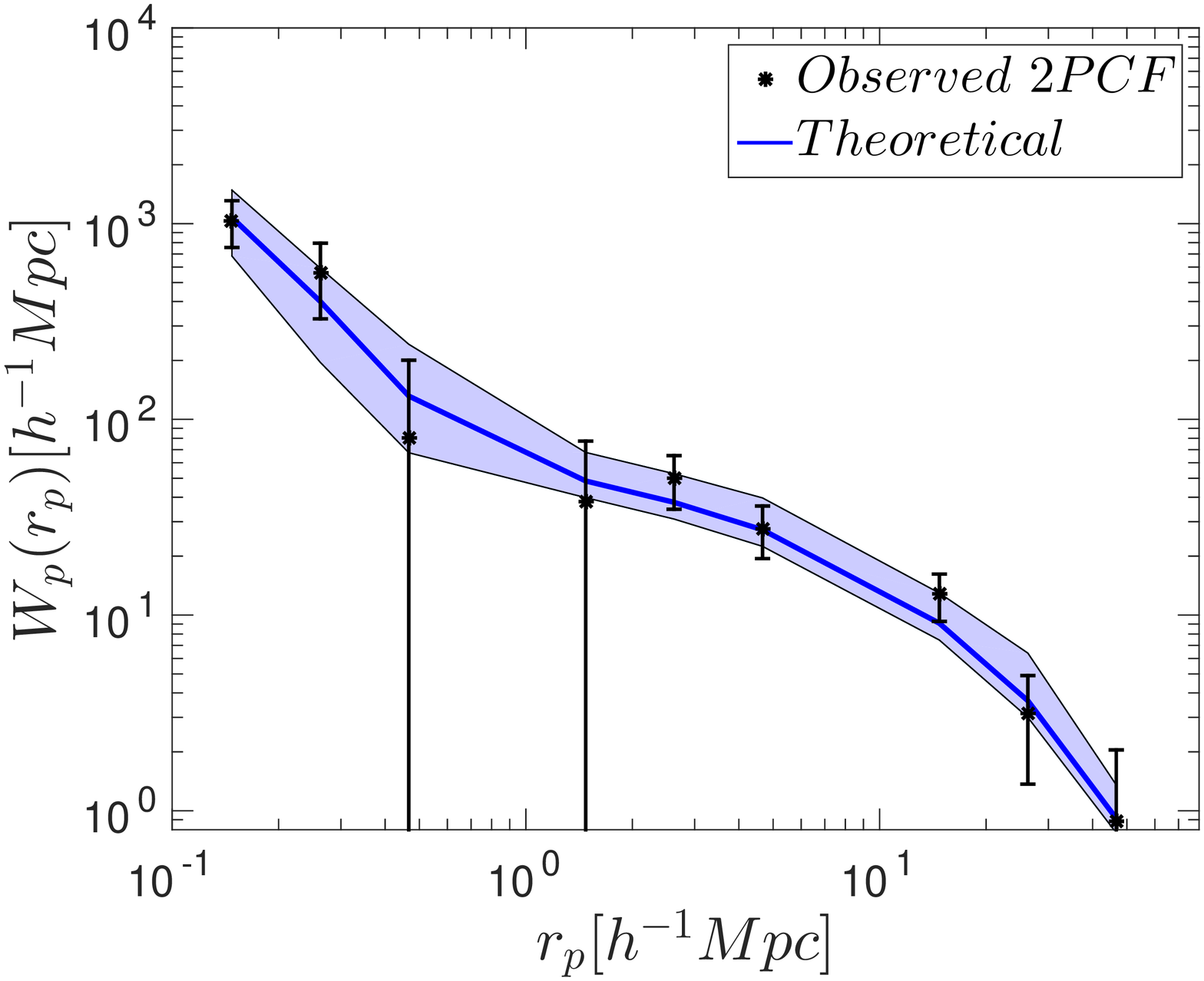}}
        \resizebox{8cm}{!}{\includegraphics{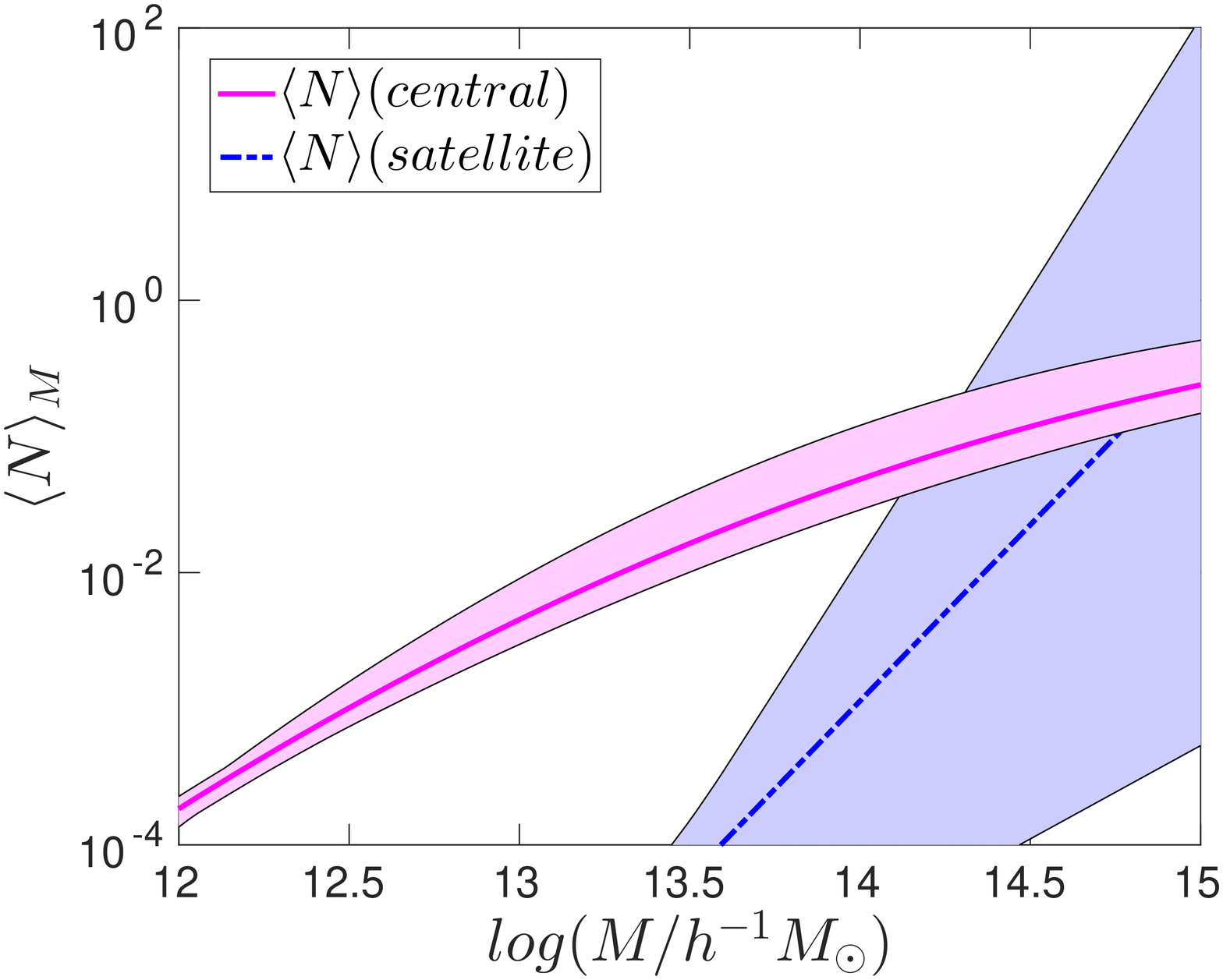}}\\
         \resizebox{8cm}{!}{\includegraphics{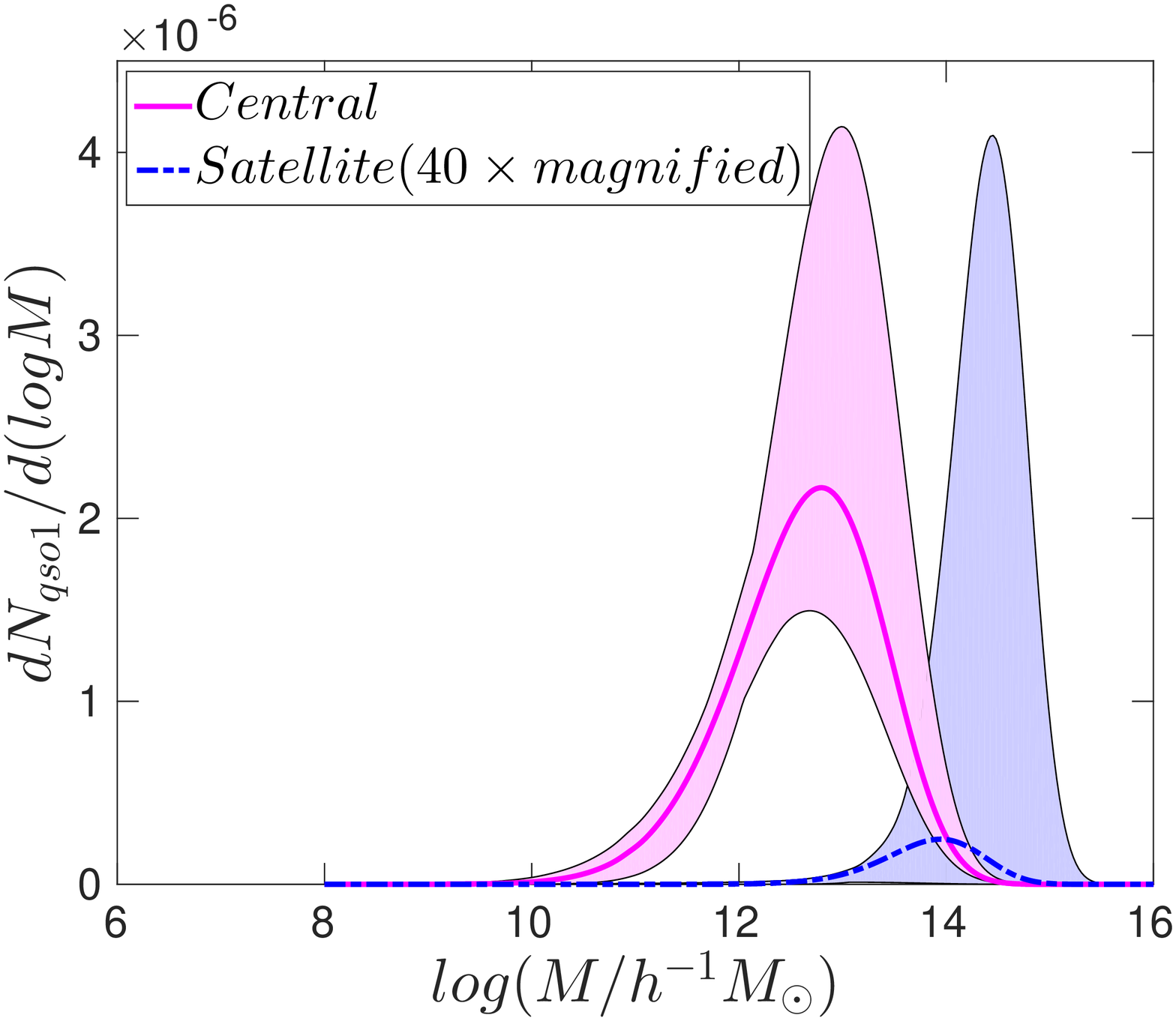}}
         \resizebox{8cm}{!}{\includegraphics{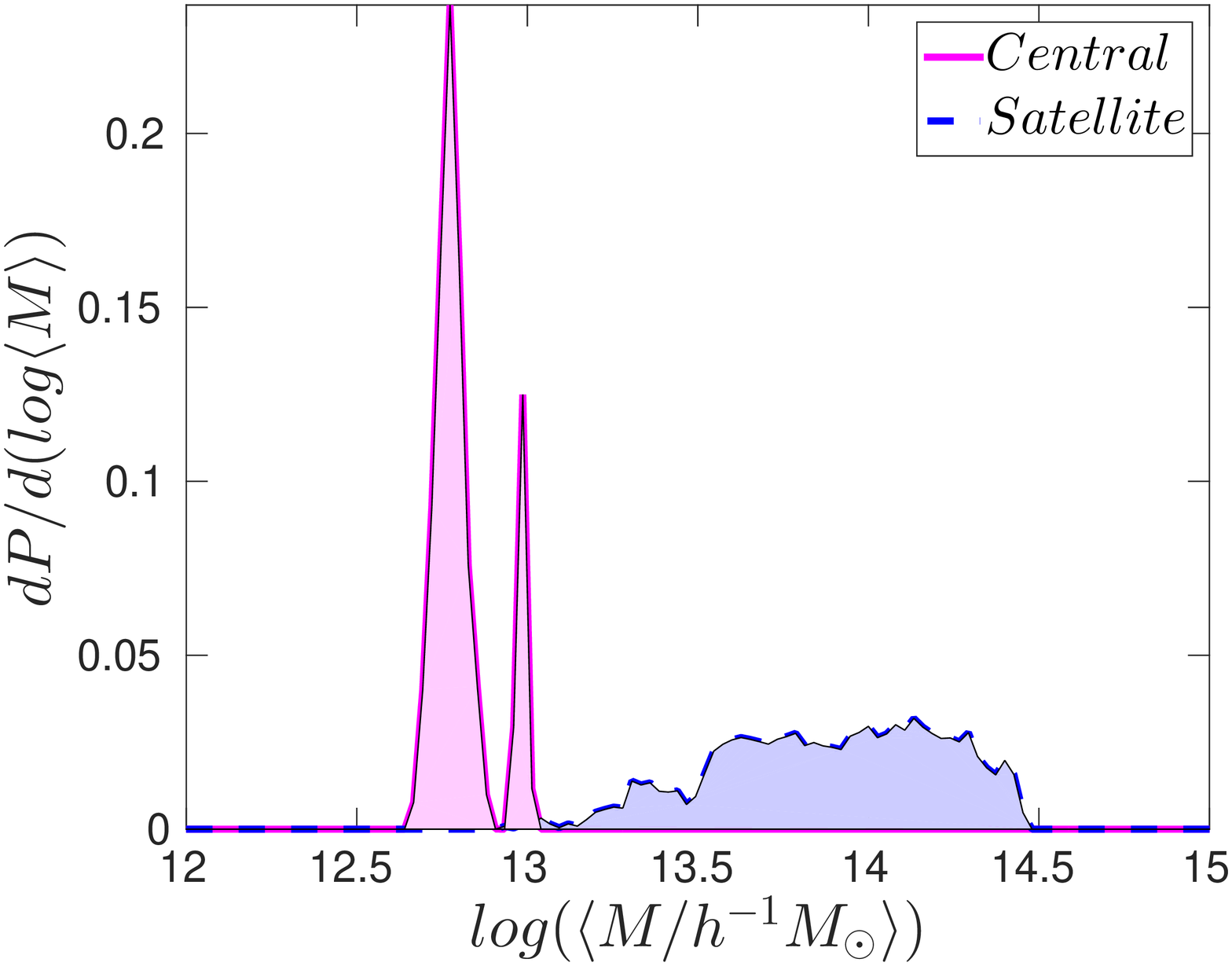}}\\
          \end{tabular}
 \caption{{\bf Top Left} : The projected 2PCF of WISE-selected unobscured quasars (median redshift $z\sim1$). The blue line is the best-fit model for the 2PCF. {\bf Top Right} : The MOF, as a function of halo mass. The magenta solid line and blue dashed line are the MOFs of the central and satellite unobscured quasars, respectively. {\bf Bottom Left} : The distribution of central (magenta solid) and satellite (blue dashed, scaled by a factor of forty for visualization purpose) unobscured quasars in dark matter halos as a function of halo mass. {\bf Bottom Right} : The probability distributions of the median mass scales of central (magenta solid) and satellite (blue dashed) quasars.}
\end{center}
\end{figure*}
\section{Datasets}

The projected 2PCF of quasars that we use in this work is constructed from the clustering sample first measured in \citet{dipompeoetal14} (henceforth referred to as D14) and updated in D16. For calculating the 2PCF, D16 uses the Landy and Szalay estimator, given as \citep{l&s93} 
\begin{equation}
\xi(\theta) = \frac{DD_{(\theta)}-2DR_{(\theta)}+RR_{(\theta)}}{RR_{(\theta)}}, 
\end{equation} 
where $DD_{(\theta)}$, $RR_{(\theta)}$ and $DR_{(\theta)}$ are defined as, respectively, the number of pairs of points separated by an angle $\theta$ in the (projected) sky, the number of pairs of points similarly separated in a random distribution, and the number of cross-pairs of points between the data and the random distributions. We refer the reader to D16 for a detailed description of the observations and datasets. Here we describe the main features of the data. 

The clustering sample has been selected from both the all-sky and the all-WISE catalogs of the WISE survey. WISE has mapped the sky in four wavebands at $3.4$\,$\mu m$, $4.6$\,$\mu m$, $12$\,$\mu m$ and $22$\,$\mu m$, referred to as $W1$, $W2$, $W3$ and $W4$. A notable feature of the D16 sample is that it satisfies the selections of both the all-sky and the all-WISE catalogs. Both obscured and unobscured quasars are observable with WISE as the hot dust in quasars is responsible for an increasing power-law spectrum in the mid-IR \citep[e.g.,][]{lacyetal04, sternetal05, donleyetal07, lacyetal13}. 

A simple color cut at $W1-W2 > 0.8$ for objects with $W2 < 15.05$ is used for selecting $225{,}303$ quasar candidates from the all-sky, all-WISE data in the region $135^{\circ}<RA<226^{\circ}$ and $1^{\circ}<DEC<54^{\circ}$. This region is chosen since it is far from the Galactic plane, and hence suffers from less foreground contamination. For details of the masking techniques, we refer the reader to D14 and D16. After the removal of various contaminants, the sample contains $175{,}911$ quasars over an area of $3422$\,deg$^{2}$. We note that given the contamination and the lack of spectroscopy, the objects we refer as quasars are actually better termed as `quasar candidates' in the truest sense.

The initial WISE-selected sample is then matched to SDSS $r$-band data, and the SDSS ``bad field'' and ``bright star'' masks are applied. The resulting sample consists of $173{,}834$ WISE-selected quasars over an area of $3387$\,deg$^{2}$. Obscured and unobscured quasars are then separated by applying an optical-IR color-cut of $r-W2 > 6$ \citep[e.g.,][]{hickoxetal07}. In addition, any WISE-selected quasars that have no SDSS counterparts in $r$-band are designated to be obscured quasars. The ultimate sample comprises $62{,}715$ obscured and $88{,}834$ unobscured quasars over an area of $3250$\,deg$^{2}$. The median redshifts of the unobscured and obscured quasars in the sample are $z \sim 1.04$ (with a standard deviation of $0.58$) and $z \sim 0.90$ (with a standard deviation of $0.54$), respectively. The entire sample of quasars covers a redshift range of $z \sim 0.1$ to $3.0$. 

To compute the number density of quasars we calculate the comoving volume in the shell between $z \sim 0.1$ and $z \sim 3.0$. We find an average number density of $2.0 \times 10^{-6}$ $(h^{-1}\,{\rm Mpc})^{-3}$ and $2.9 \times 10^{-6}$ $(h^{-1}\,{\rm Mpc})^{-3}$ for obscured and unobscured quasars, respectively. We also adopted a different method for calculating the number densities, by considering all quasars to lie in the shell between comoving radii corresponding to median $z - \sigma_{z}$ and median $z + \sigma_{z}$. We find that the results are weakly sensitive to the method of choice. While performing our HOD modeling we adopted a $15 \%$ error on our estimate of the number densities, to account for uncertainties in the redshift distributions of the quasars.

\section{Methodology}

For a given cosmology the characteristic host masses of quasars are typically obtained via bias measurements \citep[e.g.,][]{jing98,shethetal01}. However, those bias estimates do not incorporate the full halo distribution of quasars, and make no distinction between central and satellite populations. The HOD formalism, instead, allows us to extract the full distribution of the host dark matter halos of quasars from the 2PCF, which provides a more complete understanding of the relationship between quasars and their host halos. In this section, we introduce our HOD parameterization and describe the methodology by which we use the HOD to model the 2PCF.

\subsection{Halo Occupation Distribution of Quasars}
\label{sec:HOD}

The HOD of quasars is characterized by $P(N|M)$ which signifies the conditional probability that a halo of mass $M$ contains $N$ quasars combined with the spatial and velocity distributions of quasars within halos. In principle, $P(N|M)$ could be fully constructed by determining all its moments from the clustering data \citep[e.g.,][]{zhengetal07}. For our purpose of modeling the 2PCF, we need the first two moments of the distribution namely, $\langle N(M) \rangle$ and $\langle N(N-1) \rangle_{M}$ \citep{b&w02}. The HOD is assumed to be dependent only on the halo mass since the assembly bias effect is assumed to be small for the massive halos that typically host quasars \citep[e.g.,][]{bondetal91, gsw05}.

The Mean Occupation Function (MOF), or the first moment of the probability distribution, is defined as the average number of quasars in dark matter halos as a function of halo mass. We adopt a form for the MOF that consists of the sum of a softened step function for central quasars and a modified power-law for satellite quasars \citep[C12 hereafter]{chatterjeeetal12}. The MOF is then given by

\begin{equation} 
{\langle N(M)\rangle}_{\rm cen} = \frac{1}{2}\left[1+{\rm erf}\left(\frac{{\rm log} M-{\rm log} M_{\rm{min}}}{\sigma_{\rm{log M}}}\right)\right], \nonumber
\end{equation}

\begin{equation} 
{\langle N(M)\rangle}_{\rm sat} = \left(\frac{M}{M_1}\right)^{\alpha} \exp \left(-\frac{M_{\mathrm{cut}}}{M} \right), \nonumber
\end{equation}

\begin{equation}
\langle N(M)\rangle = {\langle N(M)\rangle}_{\rm cen} + {\langle N(M)\rangle}_{\rm sat},
\end{equation}
where $M_{\rm{min}}$ is the host halo mass at which the average number of quasars per halo is $0.5$, $\sigma_{\rm{log M}}$ is the transition width of the softened step function, $M_1$ refers to the mass scale at which the satellite fraction is unity, $\alpha$ is the power-law index, and $M_{\rm cut}$ is the lower mass range at which the number of satellite quasars in simulations falls off exponentially. In order to perform a more constrained fit, we excluded $M_{\rm cut}$ and conducted a four-parameter modeling of the MOF. For a given halo mass, satellite quasars in simulations are found to follow an approximate Poisson distribution \citep[e.g., C12][]{degrafetal11b}. Thus, for simplicity, we assume a Poisson distribution and a nearest integer distribution for the satellite and central quasar occupation numbers, respectively. Following R12, we assume that the occupation fractions of central and satellite quasars are uncorrelated with each other. 

This HOD model used in this work was developed from a cosmological simulation which included SMBH growth and AGN feedback \citep{dimatteoetal08}. The HOD was derived based on a black hole mass based selection \citep{degrafetal11b} and a luminosity based selection (C12) and the results showed that they differ significantly due to the scatter in the correlation between black hole mass and AGN luminosity (see C12 for discussion). Here we use the luminosity based HOD model which better represents our observed samples. In this model the central occupation asymptotically reaches its maximum value one (i.e., when every halo hosts a quasar within a luminosity threshold) while the satellite occupation increases with increasing mass. It is believed that the satellite quasars are formed mostly through secondary processes (processes that are less sensitive to the gravitational potential of the dark matter halo) such as halo mergers and hence the quasar number scales as halo mass to the first order. The parameter $M_{\rm cut}$ signifies a mass scale below which such secondary processes and satellite occupation thereof are exponentially supressed. As mentioned before, in this study we noted that our modeling stays weakly sensitive to  $M_{\rm cut}$ since it is lower than the typical halo mass scales of quasars. We thus did a four parameter fit to our model.  

We, however, note that the HOD was derived based on a low luminosity sample, due to the small volume ($34 h^{-1}$ Mpc box) of the \citet{dimatteoetal08} simulations. R12 extrapolated this HOD model to explain the clustering of bright quasars. Similarly R13 used the same HOD model to derive the host halos of X-ray selected AGN \citep{allevatoetal11}. R12 discuss the effect of theoretical bias (choice of HOD model) on derived physical parameters and the degeneracy of HOD models to current 2PCF measurements. To address this degeneracy \citet{chatterjeeetal13} developed a direct measurement technique using the MaxBCG cluster sample along with SDSS quasars which revealed that at low redshift the quasar fraction tends to increase with host halo mass supporting the C12 parameterization. Following previous work, we thus assume that the AGN HOD has a universal form which we use for interpreting the clustering measurements of WISE selected quasars. See $\S5$ for further discussion on the HOD parameterization.

To obtain the host dark matter halo population of quasars we convolve the MOF with the halo mass function (HMF). We use the HMF of \citet{jenkinsetal01} in our current model. We note that our modeling is weakly sensitive to the choice of the HMF (R12, R13). We model the radial distribution of satellite quasars within halos as a Navarro, Frenk \& White profile \citep[NFW,][]{nfw97} with the concentration-mass relation from \citet{bullocketal01},
\begin{equation}
c(M,\,z)= \frac{c_{0}}{1+z} \left( \frac{M}{M_{*}} \right)^{\beta}, 
\end{equation}
where $M_{*}$ is the nonlinear mass for collapse at $z=0$, and $\beta=-0.13$. R12 verified that the model is weakly sensitive to the choice of $c_{0}$ and hence, following R12, we adopt $c_{0}=32$. 
\begin{figure*}
\begin{center}
\begin{tabular}{c}
        \resizebox{8cm}{!}{\includegraphics{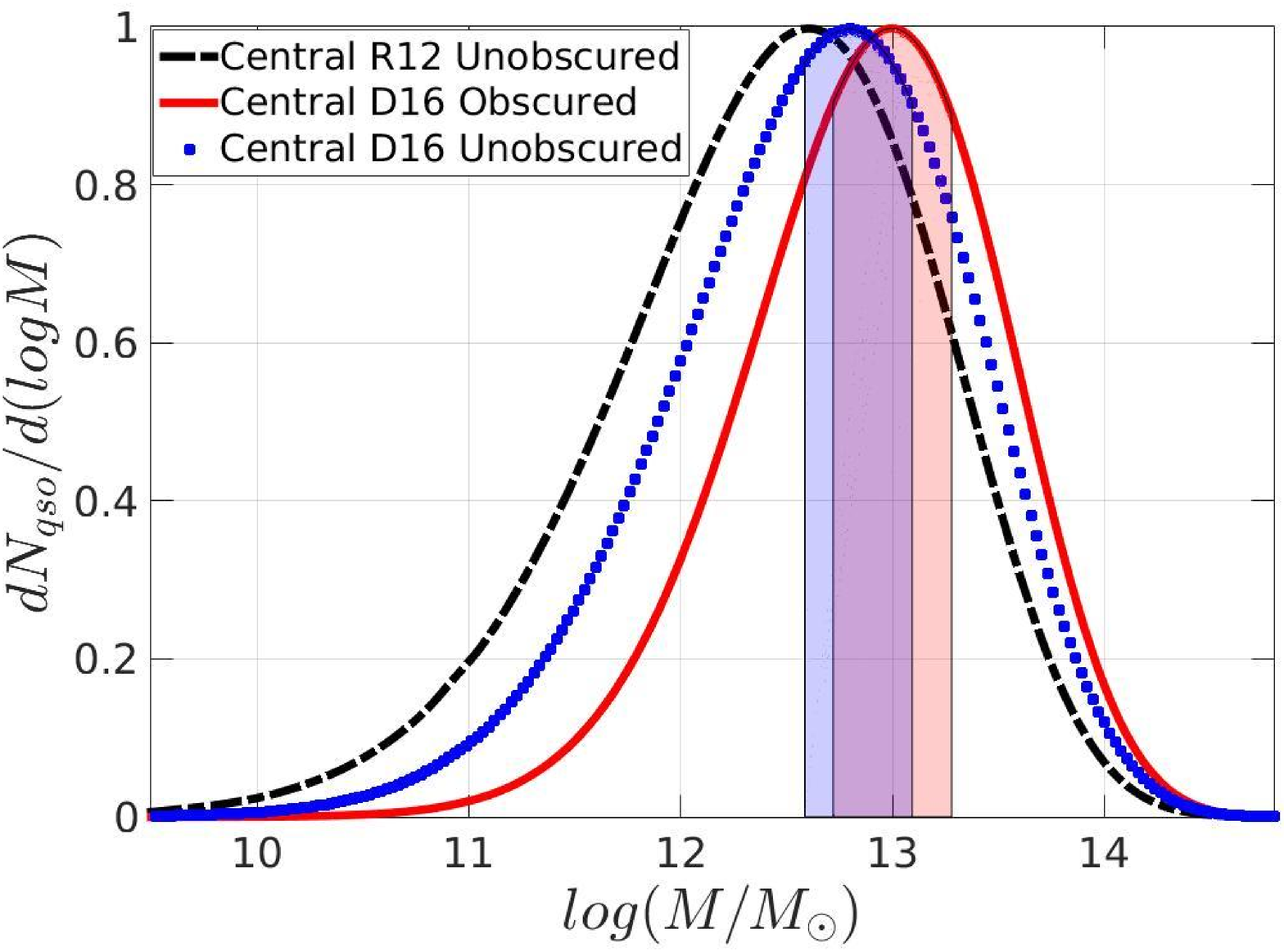}}
        \resizebox{8cm}{!}{\includegraphics{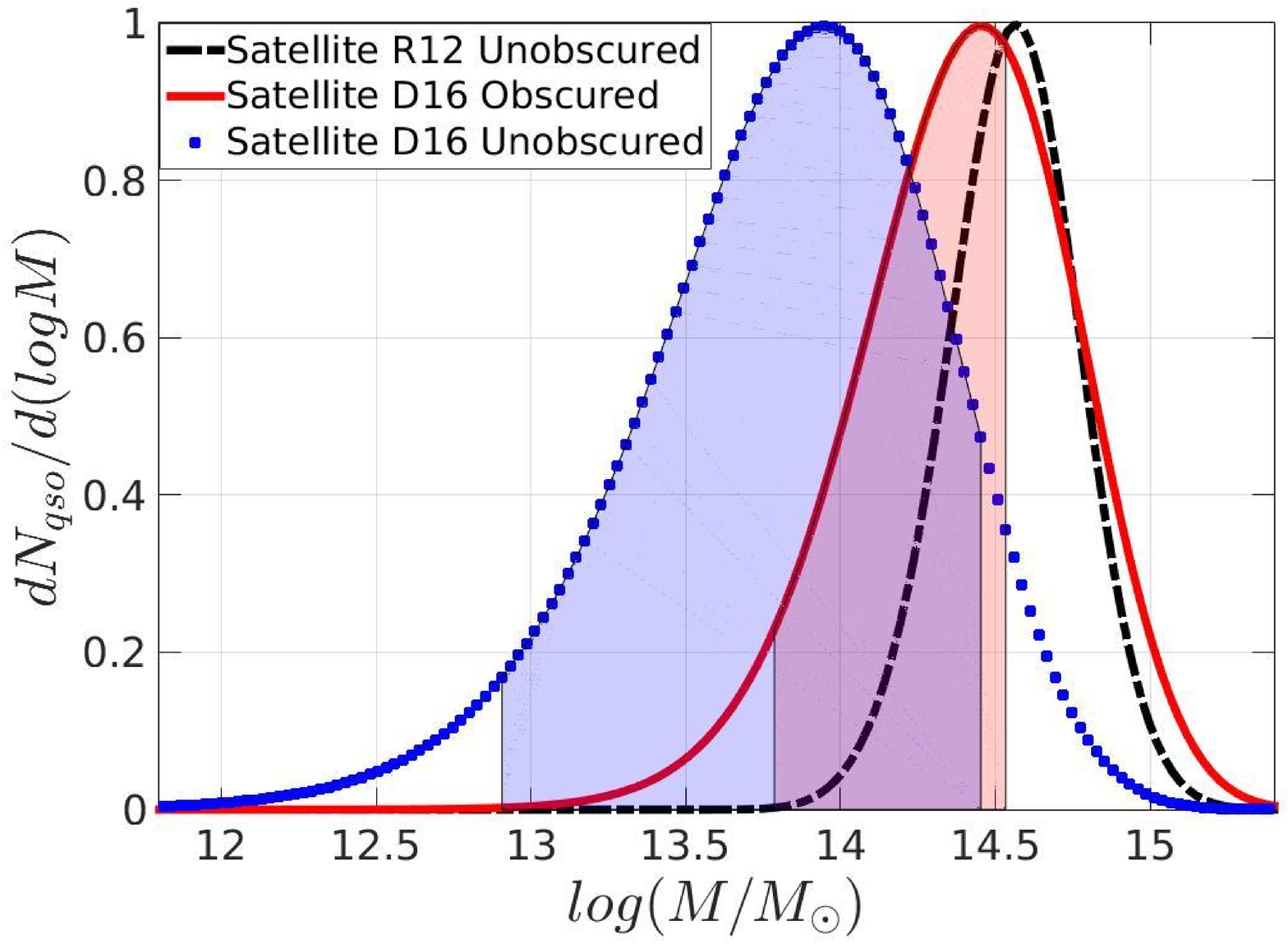}}\\
          \end{tabular}
 \caption{{\bf Comparison of host halo mass scales for three populations of quasars} : The red solid curve and blue dotted curve show the distributions for the central populations of obscured and unobscured WISE-selected quasars, respectively. The black dashed curve shows the central distribution of SDSS DR7 quasars from R12. The difference in the central mass scales between obscured D16 population and R12 is significant (1.6 $\sigma$). See Table 1 for comparison with D16 typical halo mass scales obtained from the bias measurements of quasars. {\bf Right} : Similar plot showing the distribution of satellite populations. The vertical lines with corresponding blue and red shaded regions show the one sigma errors on the medians of the central and satellite distributions. The differences in the satellite host mass scales are modest. 
}
\end{center}
\end{figure*}

\subsection{Calculation of the 2-point Correlation Function}

The quasar 2PCF, $\xi_{q}(r)$, is the excess probability of finding quasar pairs separated by a spatial distance $r$ as compared to a random distribution \citep{peebles80}. It can be decoupled into contributions from intra-halo pairs, $\xi_{1h}(r)$, and inter-halo pairs, $\xi_{2h}(r)$. The inter-halo or two-halo term can be approximated as \citep{b&w02}
\begin{equation}
\xi_{2h}(r) \approx \biggl[n_{q}^{-1} \int_{0}^{\infty} dM \frac{dn}{dM} \langle N(M)\rangle b_{h}(M)\biggr]^{2} \xi_{m}(r),
\end{equation}
where $n_{q}$ represents the number density of quasars, $dn/dM$ is the differential halo mass function, $b_{h}(M)$ is the halo bias factor, and $\xi_{m}(r)$ is the 2PCF of underlying dark matter. The term in square brackets corresponds to the quasar linear bias factor, $b_{q}$. The one-halo term can be modeled as
\begin{equation}
1+\xi_{1h}(r) \approx \frac{1}{4 \pi n_{q}^{2} r^{2}} \int_{0}^{\infty} dM \frac{dn}{dM} \left\langle N\left(N-1\right)\right\rangle _{M} \frac{dF_M}{dr},
\end{equation}
where $F_M(r)$ is the average fraction of same-halo pairs at separations $\le r$. The calculation accounts for the differences in the distributions of the central-satellite and satellite-satellite pairs \citep{zehavietal05}.

The projected 2PCF can be defined from the line-of-sight integral of the 3D correlation function $\xi(r)$ as \citep{d&p83}. 
\begin{equation}
w_p(r_p) = 2 \int_0^{r_{\parallel (\rm max)}} \xi(r)dr_{\parallel}, 
\end{equation}
where $r_{p}$ is the projected comoving transverse separation and $r_{\parallel}$ is the line of sight distance such that $r = \sqrt{r_p^2+r_{\parallel}^2}$. Eq.\ 8 is obtained by using the  box-function as the filter functions (since we are not smoothening out spatial fluctuations over our clustering scales) in the Limber approximation equation (see Eq.\ 13 of \citealt{simon07}).

Rather than working in configuration space, D16 measured the {\em angular} correlation function due to the unavailability of reliable redshifts for 
their entire sample. If $\theta$ is the angular separation of quasar pairs, corresponding to a comoving transverse separation $r_{p}$, then the number of pairs ($N(r_p)$) with separation between $r_p$ and ($r_p+dr_p$) can be obtained from the angular 2PCF ($w(\theta)$) via
\begin{equation}
N(r_p) dr_p= N \times \sigma \times \left[1+w(\theta)\right] \times 2 \pi r_p dr_p, 
\end{equation}
where $N$ is the total number of objects and $\sigma$ is the surface density of quasars.

We consider a volume of length $L$ and crossectional area $A$. If the actual number and the surface density (projected over the full size $L$) of quasars are $n$ and $\sigma$ respectively, we have: total number of quasars in that box $=$ $A \times L \times n$ $=$ $A \times \sigma$, where $\sigma$ is the surface density when all the quasars are projected on the surface perpendicular to the line of sight. Hence we can approximately write, $\sigma \approx n \times L$. Then, from Eq.\ 9 we have
\begin{equation}
N(r_p) dr_p = N \times n L \times \left[1+w(\theta)\right] \times 2 \pi r_p dr_p
\end{equation}

The pair count can also be computed from the 3D correlation function $\xi\left(\sqrt{r_p^2+r_{\parallel}^2}\right)$. We reserve $r_{\parallel}$ for denoting the line-of-sight distance between quasar pairs. The number of pairs between $r_p$ and $r_p+dr_p$ contributed from a layer chosen along the line of sight $y$ to $y+dy$ is
$$
N(r_p,y)dr_p dy = \frac{N}{L} dy \int_{-y}^{L-y} n  2 \pi r_p dr_p  \left[1+\xi\left(\sqrt{r_p^2+r_{\parallel}^2}\right)\right] dr_{\parallel}, 
$$
where $n$ is the number density of quasars, $\frac{N}{L}dy$ is number of quasars in the mentioned layer. 
Hence, the total number of pairs having projected separation $r_p$ is
\begin{equation}
N(r_p)dr_p= \int_0^L \frac{N}{L} dy \int_{-y}^{L-y} n 2 \pi r_p dr_p  \left[1+\xi\left(\sqrt{r_p^2+r_{\parallel}^2}\right)\right] dr_{\parallel}
\end{equation}


Now comparing Equations 10 and 11,
\begin{equation}
w(\theta)=\int_0^L \frac{dy}{L} \times \frac{1}{L} \int_{-y}^{L-y} \xi\left(\sqrt{r_p^2+r_{\parallel}^2}\right) dr_{\parallel}
\end{equation}
Due to large-scale homogeneity, if we assume the quasar distribution to be periodic, then $\left[(-y) \to (L-y)\right]$ integration can be equated to that over $\left[0 \to L\right]$. Then we have

$$w(\theta) = \int_0^L \frac{dy}{L} \times \frac{1}{L} \int_{0}^{L} \xi\left(\sqrt{r_p^2+r_{\parallel}^2}\right) dr_{\parallel} =  \frac{1}{L} \int_{0}^{L} \xi\left(\sqrt{r_p^2+r_{\parallel}^2}\right) dr_{\parallel}$$
therefore, using Eq. 8 we get:
$$ w(\theta)=\frac{1}{2L} \times 2\int_{0}^{L} \xi\left(\sqrt{r_p^2+r_{\parallel}^2}\right) dr_{\parallel}= \frac{1}{2L} w_p(r_p)$$
where $L=r_{\parallel}^{max}$ is the depth of the survey.

Thus, we can approximately write the angular to spatial 2PCF conversion as $w(\theta)\times 2r_{\parallel}^{max} = w_p(r_p)$. We use $r_{\parallel}^{max} = 2.88$ Gpc for obscured quasars, and $2.87$ Gpc for unobscured sample --- which are, respectively, the comoving distances corresponding to the thickness of the shell $:$ median $z \pm \sigma_z $ (which are $z \sim 0.90 \pm 0.54$ for obscured and $z \sim 1.04 \pm 0.58$ unobscured). 

We would like to note that this particular method of conversion from angular to spatial coordinates has a caveat. The given technique smoothes out the differences that can arise due to scale mixing (at a given angular separation the comoving pair separation can differ). We note that due to this conversion the large scale $w_p(r_p)$ may be slightly lower than the true value, leading to an underestimate of the halo mass of central quasars. This can also affect the satellite fraction since the slope of the correlation function will be slightly flattened this conversion. Moreover, the conversion requires projection of the quasars on an arbitrary surface of area $A$, which is not well-defined in case of a conical volume, yet constrained by the condition $A \times L \times n$ $=$ $A \times \sigma$ which needs to be satisfied. Hence the choice of average area $A$ remains somewhat flexible leading to an uncertainty in the conversion from spatial to angular scale. Our modeling works under the assumption that these effects will be minimal and can be incorporated within the uncertainties of the measurement. We discuss this further in the next section and test for its robustness.

We note that the clustering sample of D16 covers a wide range of redshift and our modeling uses halo properties that could evolve (e.g., the mass function, the halo bias factor). However, any redshift-dependence of such halo properties is not accounted for in our calculations. R12 have shown that the true HOD can be interpreted as the HOD for objects at the median redshift (within the errors on the measurement), if the 2PCF measured over a wider redshift range is statistically consistent with the actual 2PCF of the same objects at the median redshift. We adopt this interpretation for our analyses---i.e.\ we assume that the clustering evolves only weakly with redshift. Our assumption that quasar clustering is fairly constant across our redshift range of interest is supported somewhat by the measurements of bias evolution (compared to a non-evolving bias model) made by D16. We refer the reader to $\S5$, R12 and R13 for more detailed discussion regarding the limitations of this interpretation.

\section{Results}

To model the 2PCF, we use the routine developed by \citet{zhengetal07}. The code uses a Markov Chain Monte Carlo (MCMC) algorithm in the four-dimensional parameter space discussed in \S\ref{sec:HOD}. Using the underlying halo mass function from \citet{jenkinsetal01}, the code populates a virtual sky with halos, and the halos with quasars following the C12 MOF (Eqns.\ 1 and 2). Following the prescription of R13 we calculate the $\chi^{2}$ value of each point in the parameter space using the diagonal elements of the covariance matrix \citep[see][]{myersetal07a}. Each calculated $\chi^{2}$ accounts for the combined uncertainties of the 2PCF values and the number density of quasars. In our calculation, dark matter halos are defined as objects with a mean density of $200$ times that of the background density \citep[for further details about the routine see R12,][]{zhengetal07}. The MCMC contains $100{,}000$ points in the HOD parameter space, and the set of parameters with the minimum $\chi^2$ value provides the best-fit theoretical model. The error on each of the individual parameters is obtained from the procedure followed by R12. The $\chi^2$ values are arranged in ascending order starting from the minimum $\chi^2$. The envelope (of parameter values) corresponding to the $68\%$ of the values from the minimum is used to quantify the error on the best-fit parameters.

\begin{table}
\begin{center}
\caption{Halo Mass Scales of Obscured and Unobscured quasars}
\begin{tabular}{c|c|c}
\hline
\hline
\multicolumn{1}{c|}{Sample}&
\multicolumn{1}{c|}{$\log (M_{\rm cen}/\rm h^{-1}M_{\odot})$}&
\multicolumn{1}{c}{$\log (M_{\rm typical}/\rm h^{-1}M_{\odot})$}\\
\multicolumn{1}{c|}{}&
\multicolumn{1}{c|}{HOD constraints}&
\multicolumn{1}{c}{bias measurements (D16)}\\
\hline
 D16 obscured & $13.0 ^{+0.1} _{-0.2}$ & $13.0 ^{+0.14} _{-0.16}$ \\
 D16 unobscured &  $12.8 ^{+0.3} _{-0.2}$ & $12.72 ^{+0.13} _{-0.15}$\\
 R12 &  $12.61^{+0.04} _{-0.03}$  & \\ 
\hline 
\end{tabular}
\end{center}
\end{table}

\subsection{WISE-selected Obscured Quasars}

In the top-left panel of Fig.\ 1, we show our four-parameter HOD fit to the 2PCF of WISE-selected obscured quasars, at a median redshift of $z\sim 0.9$. The best-fit parameters are: $\log (M_{\rm min}/(h^{-1}M_{\odot})) = 15.26 ^{+0.92} _{-0.48}$, $\sigma_{{\rm log}M} = 1.25 ^{+0.36} _{-0.20}$, $\log (M_1/(h^{-1}M_{\odot})) = 14.56^{1.41}_{-0.02}$, and $\alpha = 3.99 ^{+0.01} _{-2.03}$. The best-fit set of parameters correspond to a reduced $\chi^2 = 1.12$ (with eight degrees of freedom). In the top-right panel of Fig.\ 1 we show the MOF from the best-fit HOD model, decomposed into its central and satellite components. The shaded regions depict the uncertainties in our estimate of the MOF.

In the bottom-left panel of Fig.\ 1, we show the host halo mass distribution of WISE-selected obscured quasars for the central and satellite populations. We have magnified the satellite distribution by a factor of 15 for visualization purposes. The peak of the satellite distribution is much lower than that of the central distribution, which is expected since the probability of finding two bright quasars in a single DM halo is minimal. The central population peaks at a halo mass of  ${\rm log}( M/(h^{-1}M_{\odot})) = 13.0 ^{+0.1} _{-0.2}$. The satellite population peaks at ${\rm log}( M/(h^{-1}M_{\odot})) = 14.4 ^{+0.1} _{-0.6}$. In the bottom-right panel of Fig.\ 1 we show the probability distributions of the median halo mass scales (obtained by multiplying the MOF with the HMF) of central and satellite quasars.

\subsection {WISE-selected Unobscured Quasars}

In the top-left panel of Fig.\ 2, we show the best-fit HOD of the observed 2PCF of WISE-selected unobscured quasars, at a median redshift of $z\sim1.04$. The best-fit parameters are : $\log (M_{\rm min}/(h^{-1}M_{\odot})) = 15.75 ^{+0.75} _{-0.94}$, $\sigma_{{\rm log}M} = 1.49 ^{+0.33} _{-0.38}$, $\log (M_1/(h^{-1}M_{\odot})) = 15.14^{2.35}_{-0.57}$ and $\alpha = 2.59^{+1.41} _{-1.27}$. The minimum $\chi^2$ of the best-fit, given $6$ degrees of freedom, corresponds to a reduced $\chi^2 = 0.44$. In the top-right panel of Fig.\ 2 we show the MOF from the best-fit HOD model, decomposed into its central and satellite components. In the bottom-left panel of Fig.\ 2 we show the host halo mass distributions of central and satellite quasars. The satellite distribution has been magnified by a factor of 40 to make it visible. The central population peaks at a DM halo mass of ${\rm log}( M/(h^{-1}M_{\odot})) = 12.8 ^{+0.3} _{-0.2}$. The satellite population peaks at ${\rm log}( M/(h^{-1}M_{\odot})) = 14.0 ^{+0.5} _{-1.0}$. In the bottom-right panel of Fig.\ 2 we show the probability distribution of the median mass scales (obtained by multiplying the MOF with the HMF).

\subsection {Comparison with Optically Selected Quasars}

The comparison between the distributions of obscured and unobscured quasars obtained from our four-parameter model and the same from R12 are shown in Fig.\ 3. In R12 the median halo masses of central and satellite quasars lie in the range $M_{\mathrm{cen}} = 4.1^{+0.3}_{-0.4} \times 10^{12} \; h^{-1} \; \mathrm{M_{\sun}}$ and $M_{\mathrm{sat}} = 3.6^{+0.8}_{-1.0}\times 10^{14} \; h^{-1} \; \mathrm{M_{\sun}}$, respectively. The central distribution of  R12 is in agreement with that of unobscured D16 quasars, $M_{\mathrm{cen}} = 6.3^{+6.2}_{-2.3} \times 10^{12} \; h^{-1} \; \mathrm{M_{\sun}}$. The median halo mass of D16 obscured quasars is higher ($M_{\mathrm{cen}} = 10.0^{+2.6}_{-3.7} \times 10^{12} \; h^{-1} \; \mathrm{M_{\sun}}$) than the halo masses of the R12 SDSS-selected unobscured population (at a level of $1.6$  $\sigma$). Our statistical significances are quoted in the sense that the lower bound on the measurement of one is consistent with the upper bound on the measurement of the other.

The typical halo mass scales of obscured and unobscured quasars as measured by D16 (via bias evolution) are shown in Table 1 for comparison with the current work.  We note that our results are consistent with D16. D16 report a slight difference in the host halo mass scales of their obscured and unobscured populations. \citet{dipompeoetal17} combines clustering and cosmic microwave background lensing measurements over a larger area and reported a difference of higher significance in the typical halo mass scales of obscured and unobscured quasars. Our results on full HOD analysis tend to favor their findings. We do not observe any significant difference in those two populations, but the differences in central host mass scales of R12 (optically selected unobscured population) and D16 obscured quasars are significant (as discussed above). Our HOD results for the D16 samples are also in agreement with that of \citet{mendezetal16} who do not find any significant differences between the clustering of obscured and unobscured populations. However, we do observe a difference in the host mass scales of R12 and D16 obscured quasars. 

This implies that typically obscured quasars tend to prefer higher mass halos than their unobscured counterparts. We would like to emphasize that although simple bias based techniques can provide constraints on host halos it is essential to exploit the full HOD prescription to truly quantify the statistical significance of the derived host halo masses from 2PCF analyses. As noted before, the HOD provides the full halo mass distribution which in turn can provide additional constraints on observed results. For example, in this work the inferred halo masses of the WISE selected obscured and unobscured quasars are similar despite being slightly different in bias-based measurements in D16. 

 \citet{allevatoetal14} use X-ray selected AGN to infer the host masses of obscured and unobscured populations. Their results show that unobscured quasars inhabit higher-mass halos compared to the obscured population. We want to emphasise that in \citet{richardsonetal13} we have done a comparison of the HOD of optically bright quasars with that of X-ray selected AGN. We do see that at similar redshifts X-ray AGN have higher mass hosts compared to optical quasars favoring the Hickox picture that was proposed for lower redshift AGN. We thus note that a comparison of X-ray-selected and IR-selected samples merits consideration. In \citet{richardsonetal13} we argue that AGN follow an evolutionary sequence from optically bright quasar phase to X-ray phase and to radio phase while their host dark matter halos grow with time. Our aim in the current paper is to examine the role of the obscured phase in this evolutionary sequence. 

At face value, there is a difference in the satellite distribution and the satellite fractions for the three populations of quasars. For R12 the satellite fraction is $f_{\mathrm{sat}}=(7.4 \pm 1.4) \times 10^{-4}$. For D16 the satellite fraction for unobscured quasars is higher, $f_{\mathrm{sat}}= 1.9 (^{+36.5} _{-1.8})\times 10^{-3}$. The D16 WISE-selected obscured population has even an order of magnitude higher satellite fraction, $f_{\mathrm{sat}}= 1.49 (^{+1.8} _{-1.48})\times 10^{-2}$, as compared to their WISE-selected unobscured counterparts. We do observe a $\sim 1 \sigma$ difference in the satellite fractions between R12 and the D16 obscured population which we consider as statistically insignificant.

R12 combined their large scale 2PCF measurements with the small scale clustering measurements of binary quasars from \citep{hennawietal06}. In the case of D16 samples we do not have such small scale measurements with WISE and hence our halo mass constraint essentially came from the two-halo term. The lack of pairs on small scales arises due to effects discussed in D14. The WISE PSF, and artifacts in WISE that have to be masked, make it difficult to measure the autocorrelation function on small scales (see D14 for details). We however note that with future surveys if we do have more information on the small scale clustering of obscured quasars, we can improve our constraints on the satellite HOD. To illustrate this fact, we repeated our HOD analysis replacing the WISE-selected quasars with a mock data set in which the error bars on the 2PCF were reduced to an optimistic $10 \%$ of their measured values. 

Fig.\ 4 shows the distribution of the host halo masses and the satellite fractions of the mock data. With these reduced error bars, the HOD formalism {\em would} show a significant difference between the HOD parameters for different populations of quasars. Although it is unlikely that the difference that we see in satellite fractions in the projected samples of D16 and R12 (based on the current observations) will solely be due to selection bias we still consider that as a possibility in explaining the observed difference. Future datasets can truly shed light on this issue. It is important to note, however, that greatly improved precision in clustering measurements for these populations would not yield a highly significant improvement in estimates of host halo masses for {\em central} quasars, compared to the current work. Recently \citet{jiangetal16} measured the small scale environments (within $100$ kpc) of low redshift seyfert samples. They found that at low redshift type 2 AGN are more strongly clustered on small scales than type 1s and that the two types have similar amplitudes on large scales. This is similar to our finding but we note that our results are drawn for a high redshift quasar sample with scales greater than $100$ kpc.


The WISE-selected quasar samples from D16 span a wide range of redshift, from $z \sim 0.1$ to $3.0$. As discussed in \S3, we interpret our derived HOD to be the true HOD at the median redshift of the D16 samples ($z \sim 1$). Essentially, we assume that any redshift evolution is incorporated within general statistical uncertainties in measurements of the clustering of quasars from D16. Were the measurement precision for the 2PCF of WISE-selected quasars to improve, such an assumption might no longer be valid. Thus, improved 2PCF measurements for WISE-selected quasars will require better modeling of the redshift evolution of the HOD in order for our formalism to yield robust conclusions regarding the mass scales of different populations of quasars. This also applies to the calculation of number densities as well as other approximations (e.g., angular to spatial conversion) adopted in this formalism. We further discuss these issues in \S5. 

As discussed in \S3.2, converting from spatial to angular scales introduces some uncertainty in our modeling. We emphasize that the effect of this uncertainty in our results is insignificant. This is further justified by the fact that the arbitrariness in conversion alters only the normalization of the projected 2PCF, by a small numerical constant of the order of unity. To test this effect of 2PCF normalization on our results, we repeated the entire analysis by increasing the normalization by a factor of two. This change produced a slight difference in the peak masses of the central quasars, which is well within the error range shown in the bottom panels of Fig.\ 1 and Fig.\ 2. The satellite halo occupation distributions remained unchanged.

\begin{figure}
\begin{center}
\begin{tabular}{c}       
         \resizebox{8cm}{!}{\includegraphics{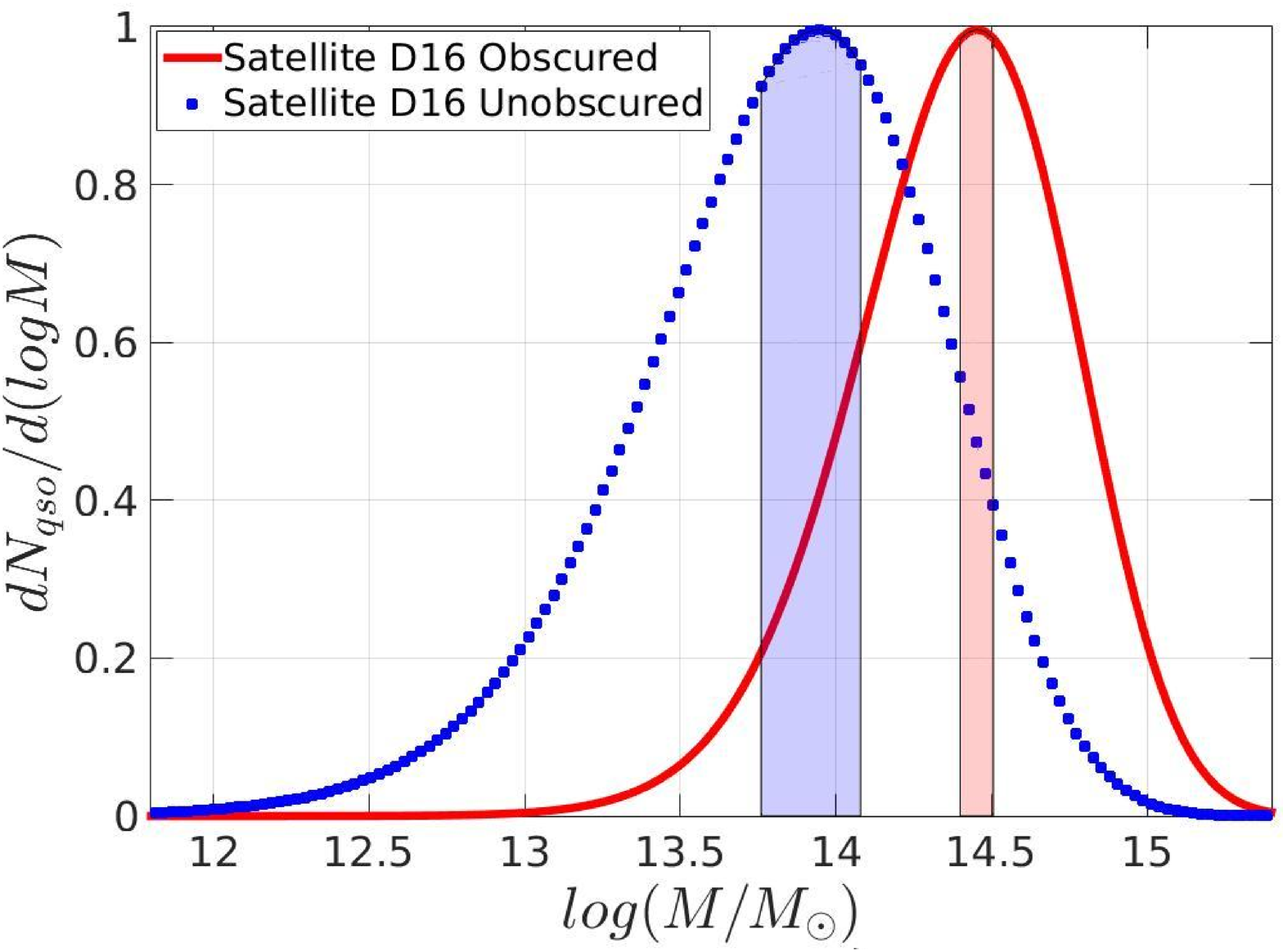}}\\
          \resizebox{8cm}{!}{\includegraphics{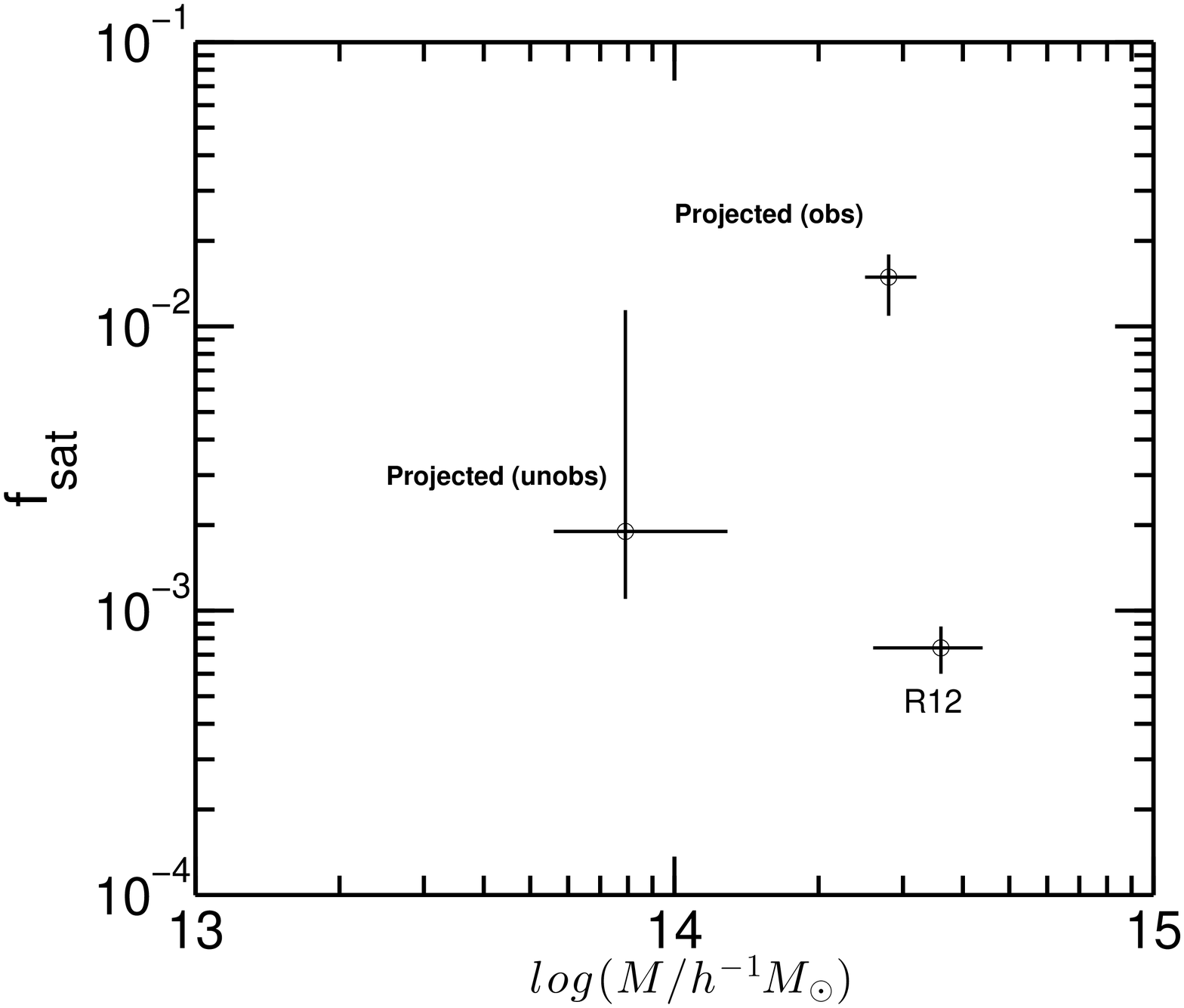}}\\
          \end{tabular}
 \caption{ The projected HOD satellite constraints with reduced errors ($\approx 10\%$ of current) on the measured 2PCF. The top panel depicts the distribution of the median mass scales while the bottom panel shows the constraints on the satellite fractions. See discussions in \S 4 and \S 5.}
\end{center}
\end{figure}

 \section{Discussion and Conclusions}

According to the simplest AGN unification theory, the central SMBH and accretion disk of a quasar are surrounded by an optically thick `dusty torus,' and the obscuration of the central broad-line-region by the torus occurs at certain inclination angles as the torus intercepts the line of sight \citep{u&p95}. In this paradigm, it would be 
expected that all three populations of quasars that we study in this paper should have identical host halo mass distributions\footnote{modulo the possibility that the samples have very different luminosities (see, e.g., D16 and \citealt{dipompeoetal17} for further discussion).}. It is possible that the simplest orientation-based unified model of AGN may be inadequate, however (see \citealt{netzer15} for a review). Within the last few decades, there have been attempts to address the nature of quasars using modified formalisms that build on the unification-through-orientation paradigm. One such formalism, the ``evolutionary theory'' of quasars \citep[e.g.,][]{sandersetal88, hopkinsetal05, hopkinsetal08, hickoxetal09} explains the origin of different types of quasars from the perspective of galaxy evolution (see \citealt{mitra16} and \citealt{dipompeoetal17} for discussion).

In a pioneering work \citet{sandersetal88} proposed that ultraluminous infrared galaxies (ULIRGs) are the initial, heavily obscured, stages of a quasar, which, after 
shedding surrounding dust, is revealed in the optical as an unobscured quasar. \citet{hopkinsetal06} proposed a merger-driven unification model in which quasar activity is triggered by galaxy mergers. Such mergers provide abundant matter, both for near-Eddington accretion on to central SMBHs and to trigger bursts of star formation in galaxies \citep[e.g.,][]{c&v00, hopkinsetal06, hopkinsetal08}. At the same time, galaxy mergers could initially enshroud central SMBHs with optically thick dust, producing an IR-bright obscured phase for quasars. As an example of evidence for this framework, \citet{chenetal13, chenetal15} find that galaxy mergers are more strongly correlated with star-formation in obscured quasars than in unobscured quasars. As accretion onto the central SMBH increases, the evolutionary paradigm suggests that, ultimately, feedback sets in \citep[e.g.,][]{c&o01, w&l03, crotonetal06, dimatteoetal05, sijackietal07, somervilleetal08, ostrikeretal10, novaketal11}, driving away the gas and dust around the central quasar. As the dust is blown away, the central quasar becomes visible in the optical and enters an unobscured phase \citep{hopkinsetal05}. 

In the context of feedback-driven evolutionary theory the satellite fractions of obscured and unobscured quasars can be different as not all of the initially obscured quasars in a halo, triggered by mergers, are expected to go to the unobscured phase; since feedback from a newly formed bright unobscured quasar could blow away gas from neighboring region. By starving the other obscured satellite AGNs, of food for accretion, it could inhibit the formation of another unobscured quasar in the same halo \citep[e.g.,][]{ostrikeretal10, chatterjeeetal12, mitra16}. This could result in a decrement of satellite fraction while going from the obscured to the unobscured phase. 

It is interesting to evaluate our work under both the ``evolutionary'' and the ``orientation'' frameworks. We note that there is a difference in the distributions of the host masses of central quasars (see Fig.\ 3) of R12 and the WISE-selected obscured population, although at a lower significance (1.6 $\sigma$). This implies that the large-scale distributions of SDSS-only-selected (R12) unobscured quasars, and the WISE-selected obscured quasars (D16), can not be fully explained by the simple unification-by-orientation scheme although that model is still consistent with our results owing to the modest statistical significance of the difference in halo masses. However, it is important to note that even in the evolutionary picture, the host halo masses of obscured and unobscured quasars can be similar if the transition time from obscured to unobscured phase is much lower than the typical halo evolution timescales. We observe that the median redshifts of R12 and D16 are $1.4$ and $0.9$ respectively. So in the evolutionary paradigm, the unobscured population of R12 should be coming from a higher redshift obscured population and the observed differences in host halo mass could as well reflect the overall redshift evolution of DM halos.

A recent work by \citet{hopkinsetal16} throws light on this issue of timescales. Using a detailed simulation in the vicinity of the SMBH, they predict a typical transition timescale of $Myr$, much smaller than the halo evolution time scales. Thus in this paradigm, one should not expect any difference in mass scales of host dark matter halos of quasars. However, satellite AGNs in a given halo might still be inhibited from going into bright unobscured phase, causing a difference in satellite fraction. So it could as well be likely that the $\sim 1.6 \sigma$ difference we found in median halo masses between R12 and D16 obscured is due to the evolution of halo mass function itself from $z \sim 1.4$ to $z \sim 0.9$ (median redshifts of the two samples). Moreover, the picture drawn by \citet{hopkinsetal16} is not purely evolutionary, it does include the orientational picture of dusty torus formation self-consistently as a result of AGN feedback, at a later stage of its evolution. So those with their torus in line-of-sight will also contribute to the obscured population, hence nullifying the halo difference even more. This is in phase with our result of statistically consistent halo mass scales of D16 obscured and unobscured populations. 

Since satellite population differences might still be present, this also suggests that in future HOD work, studying satellite fraction with more tighter constraints could be a better way, than just looking at halo mass scales, to distinguish between different stages of AGN evolution. We, however, note that there are significant differences in scenarios simulated and the one drawn in our HOD work. Starbursts, stellar outflow from galactic bulge and its coupling with the interstellar medium has a notable role to play in the duty-cycles of AGN mentioned by \citet{hopkinsetal16}. Moreover the halo masses considered in the simulation are $2 \times 10^{12} M_{\odot}$ and feedback timescales might be different in more massive halos where we have the possibility of having satellite quasars. Their study did not consider greater than $100$ pc outflow, whereas large scale (beyond galaxy scale) quasar feedback is a well-observed phenomenon.

In this work, we have shown that the C12 model derived from a cosmological hydrodynamic simulation of AGN feedback adequately explains the clustering properties of infra-red selected quasars suggesting a universal relationship between AGN activity and their host dark matter halos. We show that the HOD formalism provides more robust constraint on competing theories describing the classification of obscured and unobscured quasars and provide the very first constraint on satellite fractions of obscured quasars. In addition to that, we have for the first time proposed a technique on performing HOD modeling of angular correlation functions which are often observables in surveys where reliable spectroscopic redshifts are not available. However, we note that despite the potential statistical power of future quasar/AGN surveys and the success of HOD modeling, one of the limitations in using the HOD to model quasar/AGN clustering is the theoretical understanding of the HOD itself.

As mentioned previously the C12 model adopted in this work was constructed from a small volume cosmological simulation of low-luminosity AGN. Moreover, the model relies on a simplified subgrid-model of AGN population and AGN feedback (see C12 for discussion). We emphasize that theoretical models based on semi-analytic or numerical simulations do require extensive comparison with observations in proving their validity. The current work along with R12 and R13 do provide justification for the C12 model to be a valid HOD prescription for studying AGN co-evolution. A similar sub-grid model of AGN growth and feedback has been recently used in a large volume cosmological simulation by \citet{khandaietal15} which would further enable us to extend our HOD work to higher luminosity AGN and particularly to the population that shines as bright quasars. 

Another caveat of the current HOD parameterization lies in the redshift evolution of the HOD itself. C12 noted that the HOD parameters of the current model evolved with redshift for the low-luminosity sources, but at minor significance. There have not been any definitive studies apropos the redshift evolution of the HOD of quasars. So, in this, and in previous works, the derived HODs for quasar clustering measurements have been interpreted as the ``true'' HOD at the median redshift of the studied quasar populations (with other approximations such as number density estimates in accordance with this interpretation). This interpretation relies on the assumption that the redshift evolution of the HOD produces effects that are smaller than the statistical uncertainties of quasar/AGN 2PCF measurements. Once the statistical power of quasar/AGN clustering measurements increases, we might enter a regime where such assumptions are no longer appropriate. 

We propose to perform a study on redshift evolution of the quasar HOD with the recently run \citet{khandaietal15} simulation. We also like to refer to \citet{dipompeoetal17} for discussion on redshifts of the quasar sample. D16 did not have redshifts for all of their studied samples, notably the obscured sources. Hence even observationally it was not possible to split the sample into redshift bins for studying the redshift evolution. We plan to do a newer and deeper optical survey for carrying out the redshift analysis.  That would allow us to compare our theoretical study of the redshift evolution with that of observations. 

The understanding of quasar/AGN HOD is absolutely important in interpreting other observations \citep [e.g., Sunyaev Zeldovich: SZ effect from quasar feedback][] {ruanetal15, verdieretal16, crichtonetal16, d&c17}. Recently \citet{d&c17} showed that the uncertainty in the high halo mass tail of the mean occupation function of quasars as well as the lack of knowledge of the redshift evolution of the quasar HOD leaves the SZ detection from quasar feedback in cosmic microwave background experiments to be inconclusive. Hence the scope of the HOD work in unraveling the physical scenarios of AGN-co-evolution is extremely promising. Ultimately, a better understanding of the theoretical aspects of quasar/AGN HODs will be required in order to confidently interpret future quasar clustering measurements as well as measurements where quasars/AGN are probes of the high redshift Universe.

\section*{Acknowledgments}

We thank the referee for many useful comments which helped in improvement of the draft. KM thanks Jonathan Richardson for providing his results for comparison with the current work and also acknowledges help from Ryan Hickox for providing an earlier version of the data products based on which many of the analysis tools were tested. KM acknowledges support from the Department of Science and Technology for assistance through the KVPY fellowship. SC acknowledges support from the University Grants Commission through a start-up grant, Department of Science and Technology through the SERB-ECR grant and Presidency University through the FRPDF grant. SC is grateful to the Inter University Center for Astronomy and Astrophysics (IUCAA) for providing infra-structural and financial support along with local hospitality through the IUCAA-associateship program. ADM was partially supported by NASA through ADAP award NNX16AN48G and by the National Science Foundation through grant number 1616168. MAD was partially supported by the National Science Foundation AAG 1515404 and through NASA ADAP award NNX15AP24G.

\bibliography{KM_mybib}{}
\bibliographystyle{mn}

\end{document}